\documentclass[aps,prb,twocolumn,showpacs,amsmath,amssymb]{revtex4}
\usepackage{latexsym}
\usepackage{amsmath, amsthm, amssymb} 
\usepackage{txfonts}

\usepackage{graphicx}% Include figure files
\usepackage{dcolumn}% Align table columns on decimal point
\usepackage{bm}
\usepackage{txfonts}
\usepackage{mathrsfs}
\usepackage{feynmf}
\usepackage{comment}
\usepackage{float}

\begin{document}
\title{Quantum criticality of reconstructing Fermi surfaces in antiferromagnetic metals}

\author{Junhyun Lee}
%\email{junhyunlee@fas.harvard.edu}
\affiliation{Department of Physics, Harvard University, Cambridge MA 02138}

\author{Philipp Strack}
\email{pstrack@physics.harvard.edu}
\affiliation{Department of Physics, Harvard University, Cambridge MA 02138}

\author{Subir Sachdev}
%\email{sachdev@physics.harvard.edu}
\affiliation{Department of Physics, Harvard University, Cambridge MA 02138}

\date{\today}

\begin{abstract}
We present a functional renormalization group analysis of a quantum critical point in two-dimensional
metals involving Fermi surface reconstruction due to the onset of spin-density wave order. Its critical theory is 
controlled by a fixed point in which the order parameter and fermionic quasiparticles are strongly coupled and acquire
spectral functions with a common dynamic critical exponent. We obtain results for critical exponents and for the variation
in the quasiparticle spectral weight around the Fermi surface. Our analysis is implemented on a two-band variant 
of the spin-fermion model which will allow comparison with sign-problem-free quantum Monte Carlo simulations.
\end{abstract}

\pacs{74.40.Kb, 75.30.Fv, 75.40.Gb}

\maketitle

%\section{Introduction}

\section{Introduction}
Quantum phase transitions between two Fermi liquids, one of which spontaneously breaks translational symmetry and so reconstructs its
Fermi surface, have been of longstanding theoretical and experimental interest. Important new examples of experimental realizations
have emerged in the past few years,\cite{doiron,Kartsov2,matsuda} and so a full theoretical understanding is of some urgency. Next to immediate relevance for a class of strongly correlated electron materials, the spin-fermion model has evolved into a minimal model for itinerant lattice electrons with strong, commensurate magnetic fluctuations that are believed to destroy the Fermi liquid behavior when tuned to the critical point. How the compressible electron liquid, without Lorentz symmetry and without particle-hole symmetry, behaves when its correlations become singular, could provide some direction in the search for new universality classes beyond, for example, the better-known Gross-Neveu model of Dirac fermions which enjoys more symmetries. However, despite several
decades of theoretical work, key questions remain open especially in the important case of two spatial dimensions. 

Early theories \cite{overhauser,hertz,moriya,millis,SCS,advances} for such quantum phase transitions
focused on effective models for the quantum fluctuations of the order parameter, while
treating the Fermi surface reconstruction as an ancillary phenomenon. However, it has since become clear \cite{AC}
that such an approach is inadequate, and the Fermi surface excitations are primary actors
in the critical theory. Reference \onlinecite{metlitski10-2} postulated a critical theory for Fermi surface reconstruction, 
in which the Fermi surface excitations and the bosonic order parameter were equally important
and both acquired anomalous dimensions. These excitations were strongly coupled to each other
by a ``Yukawa'' coupling of universal strength, and their correlators scaled with a common dynamic critical exponent, $z$. 
Explicit computations were performed in the context of a $1/N$ expansion, where the physical number of fermion flavors is generalized to $N$. Taking $N$ large, one can formally reorganize Wick's theorem in powers of $1/N$ and then extrapolate results to the physical number of fermion flavors. For the hot-spot field theory at the onset of spin-density wave order,
no such critical theory appeared at the two-loop level. Indeed, it was pointed out that at higher loops \cite{lee09,metlitski10-1,metlitski10-2} there is a breakdown of the $1/N$ expansion, and so it remained unclear whether the postulated fixed point existed.

Here we address the problem of Fermi surface reconstruction at the onset of spin-density wave order
by an analysis based on a formally exact functional renormalization group (fRG) approach.\cite{berges_review02,metzner_review12} This RG approach allows a computation of correlation function as a function of a continuous cutoff scale $\Lambda$; from the ``UV'' at energies of the order of the bandwidth down to ``infrared'' excitations at and in the vicinity of the Fermi surface. Nonuniversal quantities and crossover scales  can be extracted from the same solution which also yields the critical exponents in the limit $\Lambda\rightarrow 0$. Combined with the potential to resolve the momentum (and frequency) dependence or correlators along the Fermi surface, the fRG offers much more than the field theoretic RG or conventional $\epsilon$ expansion which is typically used to extract the leading singularities only.

In this paper, we solve a set of coupled flow equations which treats the electrons on equal footing to the collective, order-parameter fluctuations. We truncate the flow equations to a set of discrete points on the Fermi surface. When projecting our correlators onto the hot spot as a function of momenta, we establish the existence of a fixed point with the scaling structure postulated in Ref.~\onlinecite{metlitski10-2}, describing the quantum phase transition between two Fermi liquids: from the metal with preserved $\textrm{SU}(2)$ spin symmetry to the metallic antiferromagnet which spontaneously breaks spin symmetry. A significant feature of our truncation 
is that it ties the parameters controlling the order parameter fluctuations to those associated with the fermion excitations, and this is important for a proper description of the scaling structure. We present numerical estimates for the critical exponents of the boson and fermion spectral functions, and for the variation in the fermionic quasiparticle residue around the Fermi surface. During our computations, we keep the shape of the Fermi surface fixed. In principle, one would have to allow for a flowing Fermi surface and consequently a flowing hot spot. In such a truncation, the singular manifold becomes a ``moving target'' and this significantly complicates the analysis.

The rest of our results are presented in Sec. \ref{result}. In Sec. \ref{model}, we introduce the recently developed two-band spin-fermion model that has the additional appealing feature that it does not suffer from the sign problem in quantum Monte Carlo simulations.\cite{berg12} In Sec. \ref{rg}, we present the functional RG setup, the truncation, and the cutoff functions. In Sec. \ref{conc}, we conclude and suggest interesting future directions resulting from this paper. 

\section{Model\label{model}}
Our computation will be carried in the context of the ``spin-fermion'' model
of antiferromagnetic fluctuations in a Fermi liquid.\cite{advances}
This involves a spin-density wave order parameter $\vec{\phi}$ at wave vector $\mathbf{K} = (\pi,\pi)$
coupled to fermions $\Psi$ moving on a square lattice. The analytic analyses have focused on the vicinity
of the ``hot spots'' on the Fermi surface: These are the eight points on the Fermi surface which can generically be
connected to each other by $\mathbf{K}$. The fermion dispersions were linearized and truncated around the hot spots.
However a complete analysis requires that we avoid the spurious singularities associated with truncated Fermi surfaces and deal
only with continuous Fermi surfaces. Here, we choose the Fermi surface configurations of a recent analysis \cite{berg12} which allowed
Monte Carlo studies without a sign problem. The present work may be seen as complementary to Ref.~\onlinecite{berg12}: Here we 
especially focus on the universality class and critical properties. 
This paves the way for an eventual comparison of our renormalization group
results with Monte Carlo. Our present method applies also to general Fermi surfaces and provides access to real-time spectral
functions which are not easily obtainable from imaginary-time Monte Carlo.

\begin{figure}[t]
\vspace*{1mm}
\includegraphics*[width=62.5mm,angle=0]{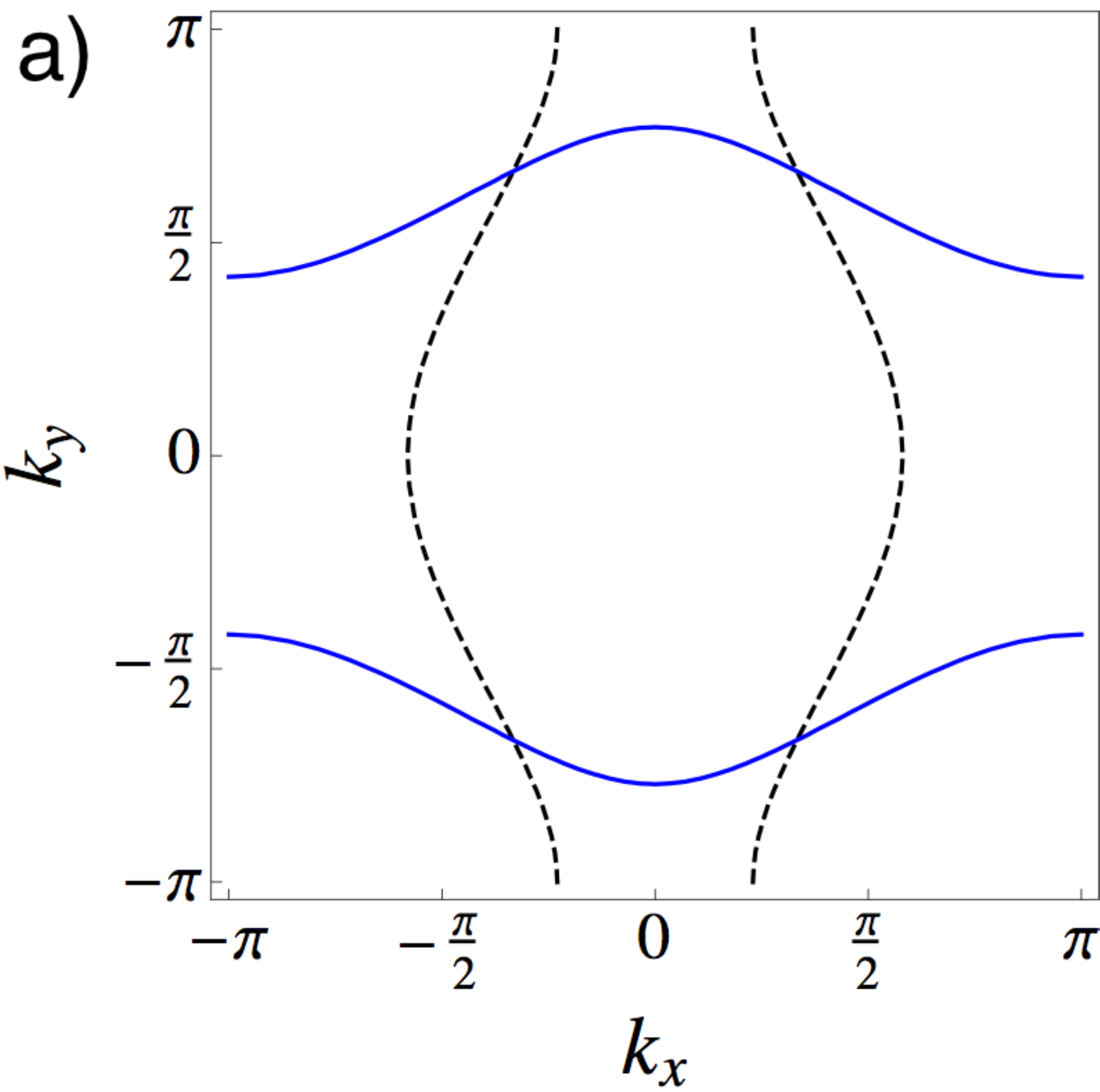}\\[5mm]
%\hspace{-5mm}
\includegraphics*[width=62.5mm,angle=0]{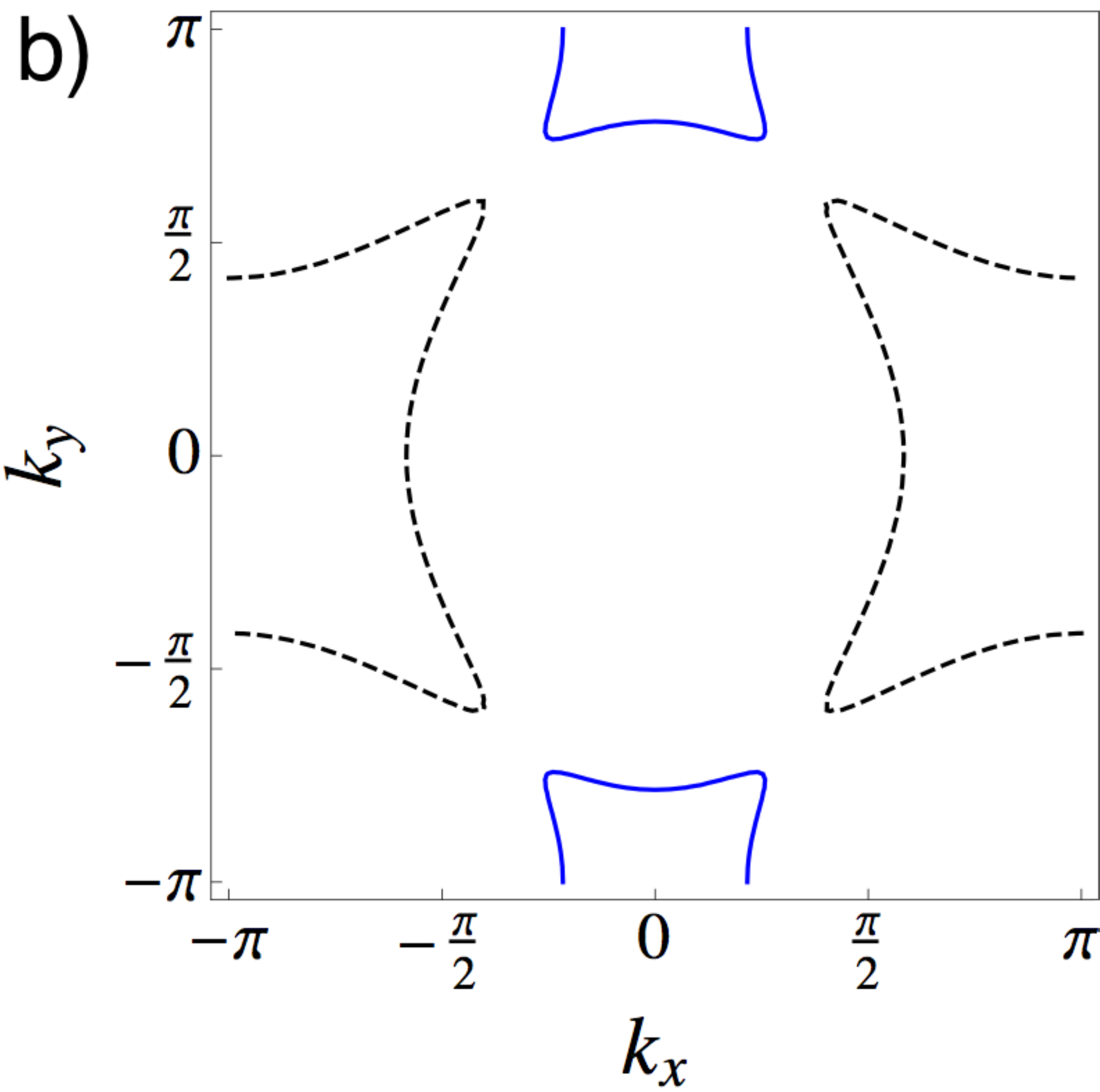}
%\vspace*{-1mm}
\caption{(Color online) Reconstructing Fermi surfaces [$ \xi_{\mathbf{k},1} = 0$, black dashed line; $\xi_{\mathbf{k},2}=0$, blue solid line 
for Eq.~(\ref{eq:fermion_dispersions})]
from the paramagnetic phase (a) to the zeros of the quasiparticle energies in the 
antiferromagnetic (SDW) phase (b). 
Gaps open at the ``hot spots,'' that is, where the Fermi surfaces of the two flavors intersect. In this paper, we focus on the 
SDW transition that is the singular point right when the Fermi surfaces reconstruct. The 
$C_4$ lattice symmetry of the original fermions is preserved.}
\label{fig:fermisurface}
\end{figure}

The model of Ref.~\onlinecite{berg12} contains fermions in two bands, or two flavors, $\Psi_{\alpha}$, $\alpha =1,2$ (although our present method
can also be applied to single band models), coupled to $\vec{\phi}$ in the effective action
\begin{align}
\Gamma^{\Lambda_{\text{UV}}}\left[\bar{\psi},\psi,\vec{\phi}\right]&=\int_k \sum_{\alpha=1,2}
\overline{\Psi}_{\alpha}(k)
\left( \begin{array}{cc}
-i k_0+\xi_{\mathbf{k},\alpha} & 0 \\
0 & -i k_0+\xi_{\mathbf{k},\alpha}
\end{array} \right)
\Psi_{\alpha}(k)\nonumber\\
&+
\int_q \frac{1}{2} \vec{\phi}(-q)\, \left(\mathbf{q}^2 + r\right) \, \vec{\phi}(q)
\label{eq:fermi_bose}\\
&+
\int_{k,q} \lambda\, \vec{\phi}(q)
\left(\overline{\Psi}_1(k+q)
{\vec{\sigma}}\,
\Psi_2(k)
+
\overline{\Psi}_2(k+q)
{\vec{\sigma}}
\Psi_1(k)
\right)\nonumber
\end{align}
where $\int_k$ represents integrals over spatial momenta $\mathbf{k}=(k_x, k_y)$ over the Brillouin zone, and over frequencies $k_0$.  
The fermion spinors are defined by 
$\overline{\Psi}_\alpha(k)= \left(\bar{\psi}_{\alpha,\uparrow}(k)\; \bar{\psi}_{\alpha,\downarrow}(k)\right)$, $\alpha=1,2$. 
We already introduce here the cutoff $\Lambda$ along which we later integrate our renormalization group flow toward $\Lambda\rightarrow 0$. 
With $\Lambda=\Lambda_{\text{UV}}$ we have the bare lattice action.
The boson quadratic term consists of the control parameter $r$ and a spatial gradient squared to account 
for spatial variations of the order parameter field $\vec{\phi}$. The quantum dynamics of $\vec{\phi}$ 
will be generated in the RG flow; putting a $q_0^2$ term into Eq.~(\ref{eq:fermi_bose}) does not change our results.
The fermion dispersions for nearest-neighbor hopping are
\begin{align}
\xi_{\mathbf{k},\alpha}=-2 t_{\alpha,x}\cos k_x - 2 t_{\alpha,y}\cos k_y -\mu_\alpha.
%\xi_{\mathbf{k},2}&=-2 t_{2,x}\cos k_x - 2 t_{2,y}\cos k_y -\mu_2\;.
\label{eq:fermion_dispersions}
\end{align}
A consistent mapping to ``physical'' fermions can be achieved with an anisotropic choice of hoppings,\cite{berg12}
$t_>=1$, $t_<=0.5$, $\mu_{\alpha} = -0.5$, and
%
%\begin{align}
$t_{1,x}=t_{>}$, 
$t_{2,x}= - t_{<}$, 
$t_{1,y}=t_{<}$, 
$t_{2,y}=-t_{>}$
yielding the Fermi surfaces shown in Fig.~\ref{fig:fermisurface}. An important distinction of this paper compared to the previous work (Refs.~\onlinecite{advances,AC,metlitski10-1,lee09,metlitski10-2,altshuler95}) is that we do not truncate the Fermi surface as patch models around hot spots. 

A mean-field analysis of Eq.~(\ref{eq:fermi_bose}) predicts 
an antiferromagnetic spin-density wave (SDW) ground state at $r=1.34$ which spontaneously breaks the spin SU(2) symmetry of 
Eq.~(\ref{eq:fermi_bose}). The Fermi surface topology ``reconstructs'' and gaps open at the hot spots, as shown in 
Fig.~\ref{fig:fermisurface}. 
On a mean-field level, the SDW transition at zero temperature of Eq.~(\ref{eq:fermi_bose}) is first order, as was also found in related single-band models for electronic antiferromagnets.\cite{altshuler95,reiss07} At present, it is not clear which effects such as fluctuations or competing instabilities could potentially drive the transitions continuous or even change the ground state. The same is true for the formation of SDWs with periods incommensurate with the underlying lattice. In the present paper, we ignore these complications and focus our attention on continuous SDW transitions at zero temperature. 

\section{Functional Renormalization Group \label{rg}}
Our RG analysis is based on the (formally exact) flow equation for the effective action $\Gamma_R^\Lambda\left[\bar{\psi},\psi,\vec{\phi}\right]$, 
the generating functional for one-particle irreducible correlation functions in the form derived by Wetterich.\cite{berges_review02,metzner_review12}
The regulator $R$ introduces a cutoff dependence into the effective action so that 
$\Gamma_{R}^{\Lambda}$ smoothly interpolates between the bare action [Eq.~(\ref{eq:fermi_bose})] at the ultraviolet scale 
$\Gamma_{R}^{\Lambda=\Lambda_{\text{UV}}} \left[\bar{\psi},\psi,\vec{\phi}\right] =\Gamma^{\Lambda_{\text{UV}}}
\left[\bar{\psi},\psi,\vec{\phi}\right]$ and the fully 
renormalized effective action in the limit 
of vanishing cutoff: $\lim_{\Lambda\rightarrow 0} \Gamma_R^{\Lambda}\left[\bar{\psi},\psi,\vec{\phi}\right] = \Gamma\left[\bar{\psi},\psi,\vec{\phi}\right]$. The Wetterich equation has a one-loop structure and in a vertex expansion the $\beta$ functions for the $n$-point correlators 
are determined by (cutoff derivatives of) one-particle irreducible one-loop diagrams with fully dressed propagators and vertices. Upon self-consistent integration of the coupled set of $\beta$ functions, contributions of arbitrary high loop order are generated. As we explain below, we truncate the effective action to the full fermion two-point function [including a fermion self-energy $\Sigma_{f}^\Lambda(k_0,\mathbf{k})$], the full bosonic two-point function [including a bosonic self-energy $\Sigma_b^\Lambda(q_0,\mathbf{q})$], and the Yukawa coupling $\lambda^\Lambda$. 

Our results are obtained from the renormalization group flow of 
the action Eq.~(\ref{eq:fermi_bose}) at the quantum-critical point ($r=0$) under the formally exact evolution 
equation \cite{metzner_review12}
\begin{align}
\frac{d}{d\Lambda} \Gamma_R^{\Lambda}\left[\chi,\bar{\chi}\right]=
\frac{1}{2}
\text{Str}
\left\{
\dot{R}^{\Lambda}
\left[
\Gamma_R^{(2)\Lambda}\left[\chi,\bar{\chi}\right]
+
R^{\Lambda}
\right]^{-1}
\right\}
%=
%\frac{1}{2}\text{Str}\,
%\left\{
%\dot{R}^{\Lambda} \partial_{R^\Lambda} \ln \left[ \Gamma_{R}^{(2)\Lambda}\left[\chi,\bar{\chi}\right]
%+
%R^{\Lambda}\right]
%\right\}
\;.
\label{eq:wetterich_equation}
\end{align}
$\Gamma_R^{(2)\Lambda}$ is the second derivative with respect to the fields defined below. $R^{\Lambda}$ is 
a matrix containing $\Lambda$-dependent cutoff functions that regularizes the infrared singularities of the fermion and boson propagators. 
The dot is shorthand notation for a scale derivative $\dot{R}^{\Lambda}=\partial_{\Lambda}R^{\Lambda}$. 
Both sides of this equation are projected onto a ``super''-field basis $\chi$, $\bar{\chi}$ containing fermionic and bosonic entries:
\begin{align}
\chi(k)
=
\left( \begin{array}{c}
\phi_x(k)\\
\phi_y(k)\\
\phi_z(k)\\
\psi_{1,\uparrow}(k)\\
\psi_{1,\downarrow}(k)\\
\bar{\psi}_{1,\uparrow}(k)\\
\bar{\psi}_{1,\downarrow}(k)\\
\psi_{2,\uparrow}(k)\\
\psi_{2,\downarrow}(k)\\
\bar{\psi}_{2,\uparrow}(k)\\
\bar{\psi}_{2,\downarrow}(k)\\
\end{array} \right)\;
\end{align}
and its conjugate-transposed $\overline{\chi}(k)$. Str is a ``super'' trace over frequency, momenta, and internal indices and installs an additional factor of $-1$ for contributions from the purely fermionic sector of the trace of Grassmann-valued matrices.
We solve Eq.~(\ref{eq:wetterich_equation}) in a vertex expansion truncating any generated vertices beyond the Yukawa vertex. The flowing fermion self-energy $\Sigma^{\Lambda}_{f}(k_0,\mathbf{k})$ and the boson self-energy $\Sigma^{\Lambda}_b(q_0,\mathbf{q})$ are parametrized in a derivative expansion keeping the Fermi surfaces fixed.

\begin{figure}[t]
\vspace*{1mm}
\includegraphics*[width=80mm,angle=0]{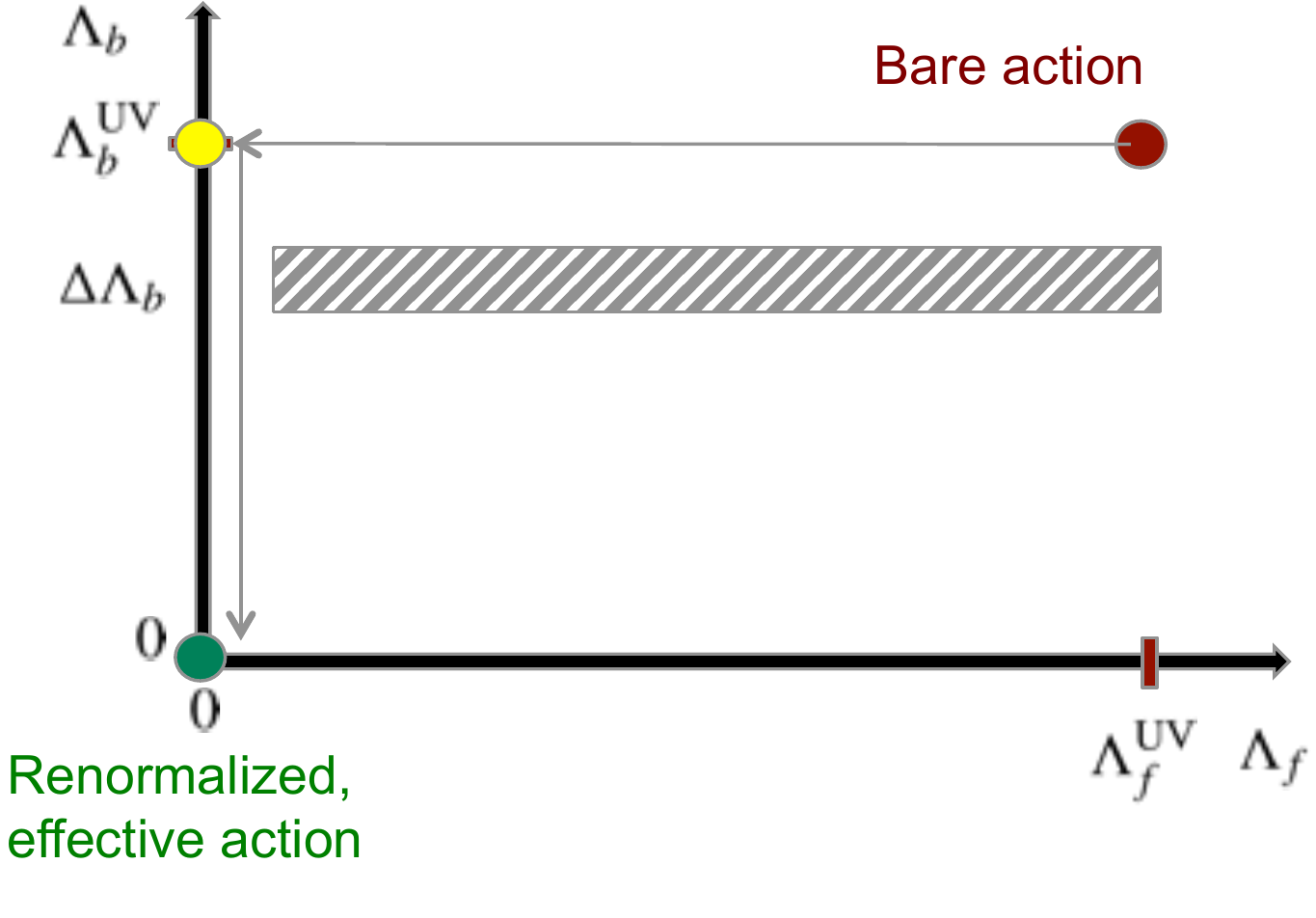}
\vspace*{-1mm}
\caption{(Color online) Illustrative flow trajectory in cutoff space. At each step $\Delta \Lambda_b$ of the integration over bosonic 
momenta along the vertical axis the entire range of fermionic momenta is swept over (gray-striped box).}
\label{fig:cutoff_space}
\end{figure}

The cutoff matrix in Eq.~(\ref{eq:wetterich_equation}) is given by
%
%\begin{widetext}
\begin{align}
R^{\Lambda}=
\text{diag}\Big(&
R^{\Lambda_b}_{b,x},
R^{\Lambda_b}_{b,y},
R^{\Lambda_b}_{b,z},
R^{\Lambda_f}_{f1,\uparrow},
R^{\Lambda_f}_{f1,\downarrow},
-R^{\Lambda_f}_{f1,\uparrow},
-R^{\Lambda_f}_{f1,\downarrow},
\nonumber\\
&
R^{\Lambda_f}_{f2,\uparrow},
R^{\Lambda_f}_{f2,\downarrow},
-R^{\Lambda_f}_{f2,\uparrow},
-R^{\Lambda_f}_{f2,\downarrow}
\Big)\;,
 \label{eq:cutoffmatrix}
\end{align}
%\end{widetext}
%
where one is, in principle, free to choose the fermion and boson cutoff scales $\Lambda_b$ and 
$\Lambda_f$ and associated regulator functions $R_{b,f}$ independently.\cite{schutz05,drukier12} The corresponding 
``flow trajectories'' in cutoff space (in the plane of Fig.~\ref{fig:cutoff_space}) from the bare action (red dot) 
to renormalized, effective action (green dot) 
will be different. We choose the trajectory along the arrows illustrated in Fig.~\ref{fig:cutoff_space}; that is, we take
$\Lambda_f\rightarrow 0$ and $R_f\rightarrow0$ before integrating out order parameter fluctuations which are excluded for momenta smaller than $\Lambda_b$. The fermions are, however, not discarded as in the Hertz theory,\cite{hertz} but coupled 
self-consistently into the flow for all $\Lambda\in\{\Lambda^{\text{UV}}_b,0\}$, thereby imposing important boundary conditions for the integration of order parameter fluctuations down the vertical axis in Fig.~\ref{fig:cutoff_space}. This makes the flow nonlocal in the cutoff scale in that the purely fermionic contractions with Yukawa vertices are treated as a total scale derivative that also acts on the self-energy on the internal lines and the Yukawa vertices. This is similar in spirit to the Katanin scheme, where this can be shown to lead to the inclusion of higher $n$-point vertices in the flow.\cite{salmhofer04}

For the bosons, we use a Litim cutoff for momenta,
\begin{align}
R^{\Lambda_b}_{b,x}=R^{\Lambda_b}_{b,y}=R^{\Lambda_b}_{b,z}=R^\Lambda_{b}&=A^{\Lambda}_{b}\left( -\mathbf{q}^2
 + \Lambda^2\right) \theta\left(\Lambda^2 - \mathbf{q}^2\right)
%R^{\Lambda_f}_{f1,\uparrow} = R^{\Lambda_f}_{f1,\downarrow} =
%R^{\Lambda_f}_{f2,\uparrow} = R^{\Lambda_f}_{f2,\downarrow} &=0
\;,
\label{eq:cutoffs}
\end{align}
where $A^{\Lambda}_{b}$ is bosonic momentum renormalization factor to be specified below. 
In the following, we set $\Lambda^b=\Lambda$. The fermionic entries in Eq.~(\ref{eq:cutoffmatrix}) 
are zero.

The fermionic matrix elements of the generalized 
matrix propagator $\left[
\Gamma_R^{(2)\Lambda}\left[\chi,\bar{\chi}\right]
+
R^{\Lambda}
\right]^{-1}$
 occurring in Eq.~(\ref{eq:wetterich_equation}) become
\begin{align}
G^{\Lambda}_{f1,\sigma}(k)&=-\langle \psi_{1,\sigma}(k)\bar{\psi}_{1,\sigma}(k)\rangle^{R}
\nonumber\\
&=-\left[\frac{ \overrightarrow{\delta}}{\delta \overline{\chi}(k_1)} 
\Gamma_R^{\Lambda}[\chi,\overline{\chi}]
\frac{ \overleftarrow{\delta}}{\delta \chi (k_2)} 
+
R^{\Lambda}
\right]_{ \substack{ f1,\sigma \\ \overline{\chi}=\chi = 0 \\ k_1=k_2=k }}^{-1}
\nonumber\\
&=
\frac{-1}{- ik_0+\xi_{\mathbf{k},1} + \Sigma^{\Lambda}_{f1}(k_0,\mathbf{k})}\;,
\label{eq:ferm_propagator}
\end{align}
and analogously for the other flavor and spin components. 

The explicitly cutoff-dependent boson spin fluctuation propagators are
\begin{align}
D^{R}(q)\equiv D^{R}_{x}(q)&=-\langle \phi_{x}(q) \phi_x(-q) \rangle^{R} 
\nonumber\\
&= -\left[\frac{ \overrightarrow{\delta}}{\delta \overline{\chi}(q_1)} 
\Gamma_R^{\Lambda}[\chi,\overline{\chi}]
\frac{ \overleftarrow{\delta}}{\delta \chi (q_2)} 
+
R^{\Lambda}
\right]_{\substack{ b,x\\ \overline{\chi}=\chi = 0 \\ q_1=q_2=q}}^{-1}
\nonumber\\
&=
\frac{-1}{ \mathbf{q}^2 + r + \Sigma^{\Lambda}_{b}(q_0,\mathbf{q})+ R^{\Lambda}_{b}}
\nonumber\\
&=
\left\{ \begin{array}{cl} \frac{-1}{ \mathbf{q}^2 + r + \Sigma^{\Lambda}_{b}(q_0,\mathbf{q}) } & 
 |\mathbf{q}| > \Lambda
\\[2mm]
 \frac{-1}{\Lambda^2 + r + \Sigma^{\Lambda}_{b}(q_0,\Lambda)} & |\mathbf{q}| < \Lambda 
 \end{array} \right.
 \;,
\label{eq:bose_prop}
\end{align}
and analogously for the other spin projections $y$, $z$. The functional derivatives are evaluated at zero fields here, 
as we approach the QCP from the paramagnetic phase. 

\begin{figure}[b]
\vspace*{0mm}
\includegraphics*[width=50mm,angle=0]{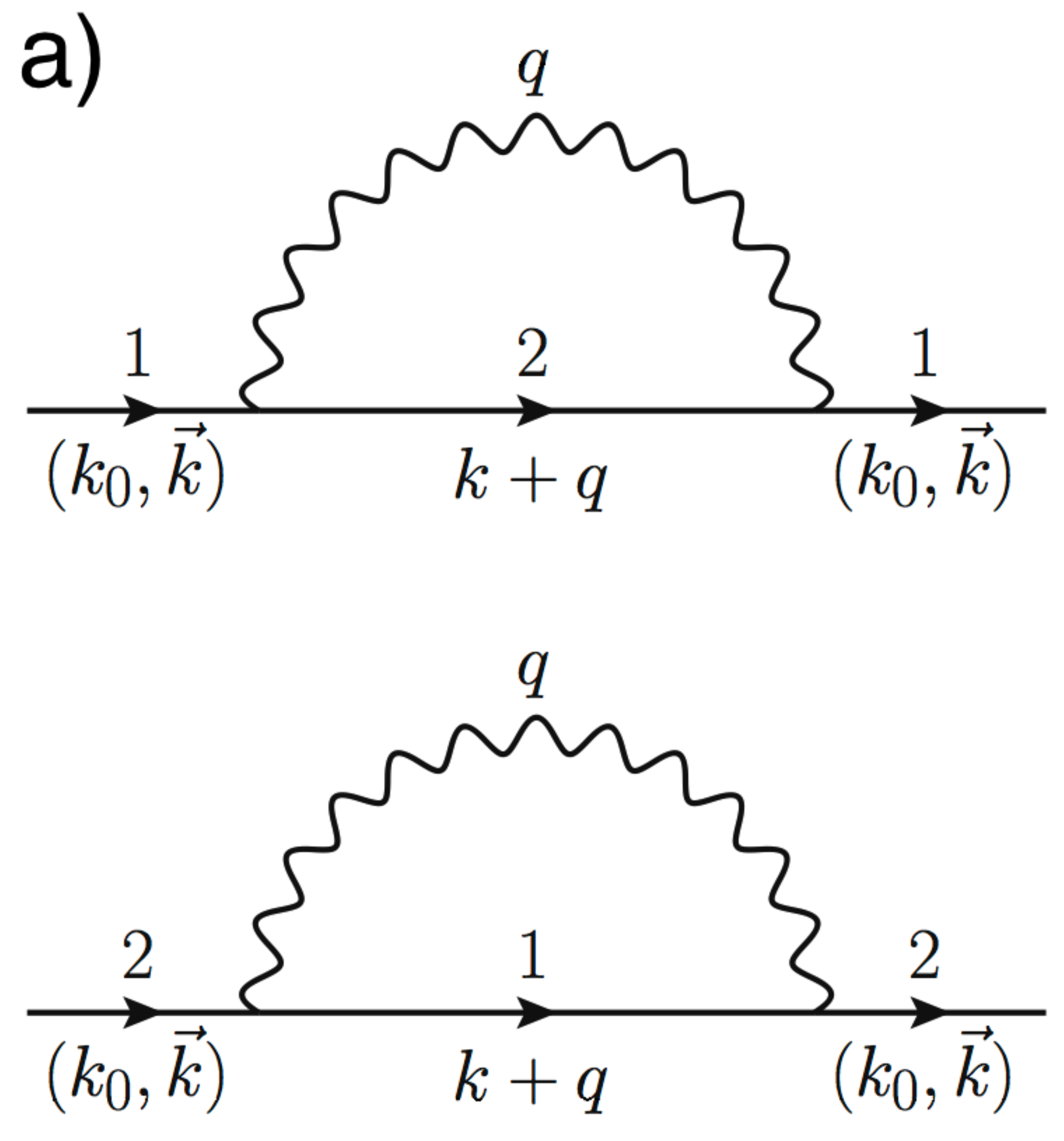}
\includegraphics*[width=50mm,angle=0]{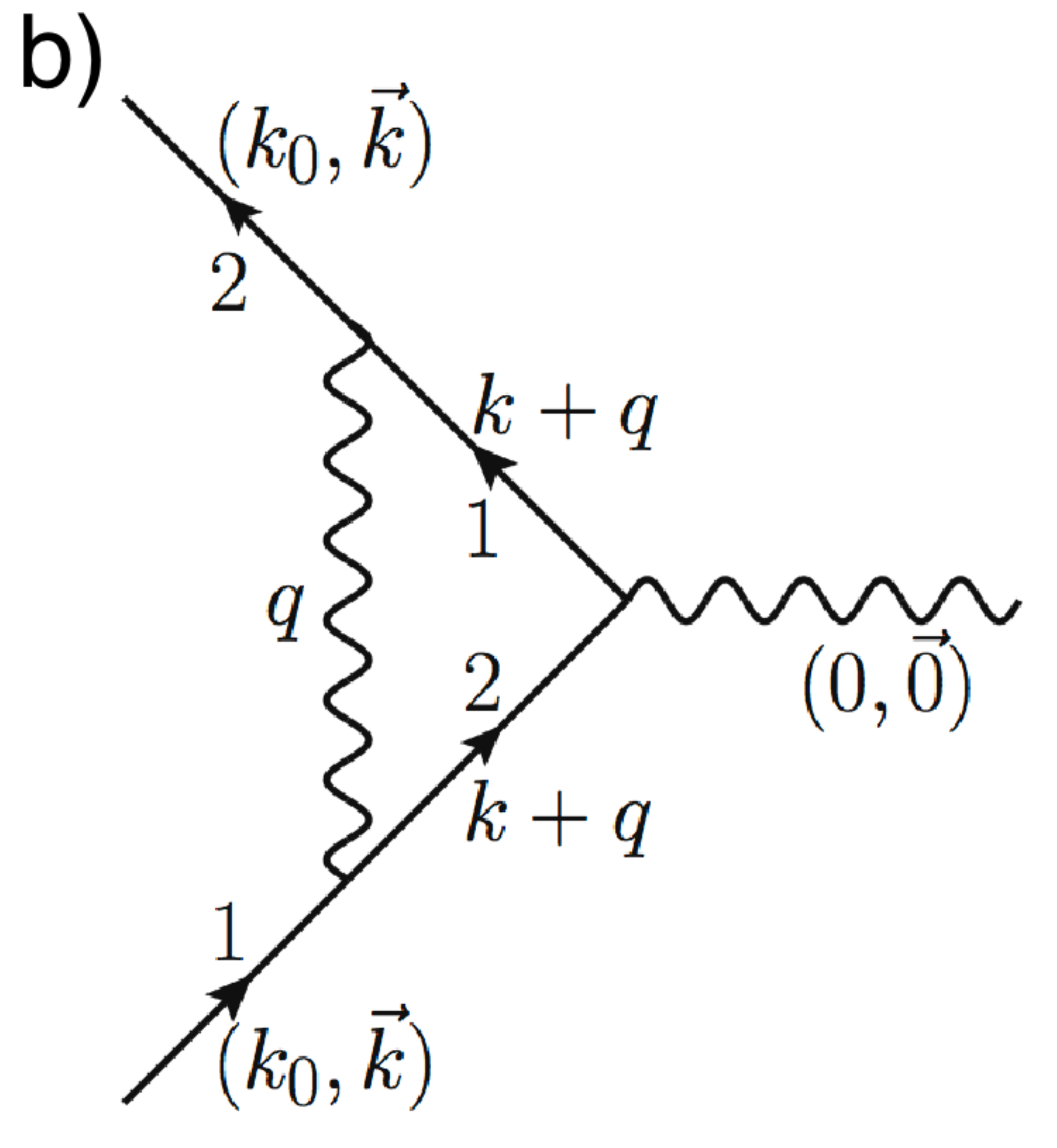}
%\vspace*{0mm}
\caption{Diagrammatic representation of the flow equation for the fermion self-energy 
$\Sigma^\Lambda_{f}(k_0,\mathbf{k})$ (a) and the Yukawa coupling (b). Straight lines 
denote Fermi propagators of flavors 1 and 2; wiggly line boson propagators are endowed with a regulator $R^\Lambda$. Intersections of wiggly with straight lines 
represent the Yukawa coupling. The cutoff derivative with respect to $R^\Lambda$ is implicit. All propagators and vertices are ``dressed'' self-consistently and are functions of $\Lambda$.}
\label{fig:flow_ferm}
\end{figure}

The flow equation for the fermion self-energy [depicted diagrammatically in Fig.~\ref{fig:flow_ferm} (a)] is
\begin{align}
\partial_{\Lambda}\Sigma_{f1}^{\Lambda}[k_0,\mathbf{k}] = 
3\left(\lambda^{\Lambda}\right)^2 \int_{q,R} 
G^\Lambda_{f2}(k+q) D^{R}_b(q)\;,
\label{eq:flow_ferm}
\end{align}
and similarly for flavor $2$ upon interchanging $1\leftrightarrow 2$.
We use a shorthand notation encapsulating frequency, momentum integrations, and a cutoff derivative 
with respect to the bosonic cutoff function:
%
%\begin{align}
$\int_{q,R_b}=\int \frac{d q_0}{2\pi} \int \frac{ d^2 \mathbf{q}}{(2\pi)^2} 
 \left[
 -
\dot{R}^{\Lambda}_{b}
\partial_{R_{b}^{\Lambda}}\right]$.
%\label{eq:cutoff_deriv}
%\end{align}
%
%

The prefactors and signs of the flow equations are computed by comparing 
coefficients between the left- and right-hand sides of Eq.~(\ref{eq:wetterich_equation}) as outlined in Sec.II   
of Ref.~\onlinecite{gies02}. The 11$\times$11 Grassmann-valued (super-) matrices are evaluated using the GrassmannOps.m package in
Mathematica. How to take a supertrace can be found in Ref.~\onlinecite{wegner98}.

\begin{figure}[t]
\vspace*{0mm}
\includegraphics*[width=65mm,angle=0]{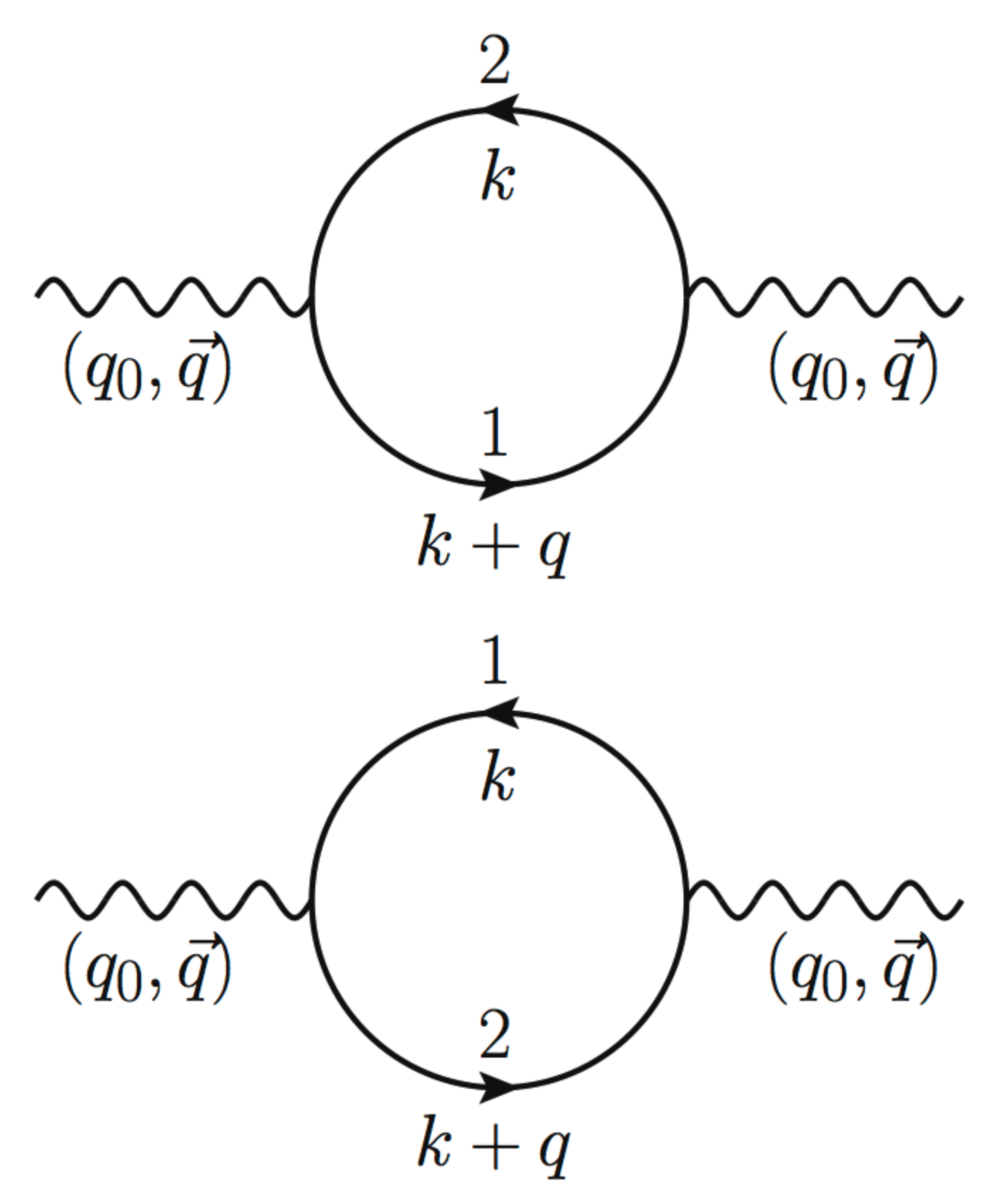}
%\vspace*{0mm}
\caption{Particle-hole bubbles used for the flow of the boson  
self-energy in Eq.~(\ref{eq:Sigma_b}). 
All propagators and vertices are ``dressed'' self-consistently and depend on $\Lambda$.}
\label{fig:boson_self}
\end{figure}

The boson self-energy is determined self-consistently from the particle-hole bubble (Fig.~\ref{fig:boson_self}) 
at all stages of the flow:
\begin{align}
\Sigma^{\Lambda}_b(q_0,\mathbf{q})
&=-\left(\Pi^{\Lambda}(q_0,\mathbf{q})-\Pi^{\Lambda}(0,\mathbf{0})\right)
\label{eq:flow_bos}\\
&= 2 \left(\lambda^\Lambda\right)^2 \int_k 
\Big[
\left(G_{f1}^{\Lambda}(k+q) - G_{f1}^{\Lambda}(k)\right)G_{f2}^{\Lambda}(k) 
\nonumber\\ 
&\hspace{18mm}+
G_{f1}^{\Lambda}(k)\left(G_{f2}^{\Lambda}(k+q) -G_{f2}^{\Lambda}(k)\right) 
\Big]. \nonumber
\label{eq:Sigma_b}
\end{align}
The following ansatz captures the leading frequency and momentum-dependence of the particle-hole 
bubble:
\begin{align}
\Sigma^{\Lambda}_b(q_0,\mathbf{q}) = Z_b^\Lambda |q_0| + (A_b^\Lambda -1)\mathbf{q}^2\;.
\end{align}
At the yellow dot in Fig.~\ref{fig:cutoff_space}, the Fermi propagators are still Fermi-liquid like 
($\Sigma^{\Lambda_{\rm UV}}_{f\alpha}=0$) because 
we have not yet integrated out any order parameter fluctuations which, by Fig.~\ref{fig:flow_ferm} (a), 
generate a finite fermion self-energy. At that point, the coefficients $Z_b^{\Lambda_{\rm{UV}}}$, 
$A_b^{\Lambda_{\rm{UV}}}$ take finite numerical values. At all stages of the flow,
when integrating the flow down the vertical axis of Fig.~\ref{fig:cutoff_space}, the bosonic $Z$ and $A$ factor are determined self-consistently according to the prescription
\begin{align}
Z_b^\Lambda&=-\frac{\Pi^\Lambda(q_0,\mathbf{0})-\Pi^\Lambda(0,\mathbf{0})}{q_0}\Big|_{q_0=\Lambda} ,\nonumber\\
A_b^\Lambda&=1-\frac{\Pi^\Lambda(0,\mathbf{q})-\Pi^\Lambda(0,\mathbf{0})}{\mathbf{q}^2}\Big|_{q_x=\Lambda, q_y = 0 }\;.
\label{eq:ZA_b}
\end{align}
This allows them to pick up potentially singular renormalizations during the flow.
The boson momentum factor is isotropic in momentum space; interchanging $q_x\leftrightarrow q_y$ delivers the same value 
for $A^\Lambda_b$.

The flow equation as per Fig.~\ref{fig:flow_ferm} (b)
for the Yukawa coupling is
\begin{align}
\partial_{\Lambda} \lambda^\Lambda = 
-\left(\lambda^{\Lambda}\right)^3 \int_{q,R} 
G^\Lambda_{f1}(k+q)G^\Lambda_{f2}(k+q) D^{R}_b(q)\Big |_{k_0=0,\, \mathbf{k}=\mathbf{k}_{\text{HS}}}\;.
\label{eq:flow_yuk}
\end{align}

The explicit expressions of the flow equations and the numerical parameter used are given in the Appendix. 

\section{Results\label{result}} 
We now describe the key results obtained from a solution of the flow equations.
(i) We find an infrared strong-coupling fixed point for the Yukawa-coupling $\lambda^\Lambda$ which governs the RG flow of the coupled Fermi-Bose action 
down to the lowest scales $\Lambda\rightarrow 0$. This induces scaling relations among the anomalous exponents for the Fermi velocity, the quasiparticle weight, and the Yukawa vertex. (ii) Both the quasiparticle weight and the Fermi velocity vanish as a power law when scaling the 
momenta toward the hot spot; the Fermi velocity slower than the quasiparticle weight. (iii) The (quantum) dynamical scaling of the 
electronic single-particle and collective spin fluctuations follows from an emergent dynamical exponent, attaining the same (fractional) value for both fermions and bosons.

The centerpiece of our analysis is the flow equation for the Yukawa coupling:
\begin{align}
\Lambda \partial_\Lambda \tilde{\lambda}^{\Lambda}
=
\left(
\frac{1}{4} \left(
\eta_{Z_{f1}}+
\eta_{Z_{f2}}+
\eta_{A_{f1}}+
\eta_{A_{f2}}
\right)-
\eta_{\rm{yuk}}-
\frac{1}{2}
\right) 
\tilde{\lambda}^\Lambda\;,
\label{eq:flow_yuk_rescaled}
\end{align}
where
%
%\begin{align}
$\left(\tilde{\lambda}^{\Lambda}\right)^2
=
\left(\lambda^\Lambda\right)^2/(\Lambda \sqrt{Z^\Lambda_{f1}Z^\Lambda_{f2}} \sqrt{A^\Lambda_{f1}A^\Lambda_{f2}})$
%=
%\frac{\left(\lambda^\Lambda\right)^2}{\Lambda Z^\Lambda_{f1}Z^\Lambda_{f2} 
%\sqrt{ |\upsilonup^{\Lambda}_{f1}| |\upsilonup^{\Lambda}_{f1}|}}\;,\quad
%\label{eq:yuk_rescaled}
%\end{align}
%
is rescaled by the frequency ($Z^\Lambda_{f1}$) and momentum ($A^\Lambda_{f1}$) derivatives of the fermion self-energy  
generated under the RG flow as per Fig.~\ref{fig:flow_ferm} (a). The power-law divergences as well as all other nonuniversal contributions 
 to the flow of the two fermion self-energy factors and the Yukawa coupling itself are absorbed into the anomalous exponents:
\begin{align}
\eta_{Z_{f1}}=-\frac{ d \ln Z^\Lambda_{f1}}{d \ln \Lambda}\;,\quad
\eta_{A_{f1}}=-\frac{ d \ln A^\Lambda_{f1}}{d \ln \Lambda}\;,\quad
 \eta_{\rm{yuk}}=-\frac{ d \ln \lambda^{\Lambda}}{d \ln \Lambda}\;.
\label{eq:ferm_anomalous}
\end{align}
$\eta_{\text{yuk}}$ is driven by the direct contribution to the flow of $\lambda^\Lambda$ exhibited in Fig.~\ref{fig:flow_ferm} (b).
All couplings are projected to zero fermionic frequency, a discrete set of fermionic momenta on the Fermi surfaces, and zero bosonic frequency and momenta. This is where the most singular renormalizations occur.

Specifically, the inverse quasi-particle weight is computed from the flowing self-energy by \cite{honerkamp03}
\begin{align}
Z^{\Lambda}_{f1}
%\equiv Z^{\Lambda}_{f1}[k_0 = 0, \mathbf{k}=\mathbf{k}_{\text{HS}}]
=1-\frac{\partial}{\partial i k_0} \Sigma^{\Lambda}_{f1}(k_0,\mathbf{k})|_{k_0=0, \mathbf{k}=\mathbf{k}_{\text{F}}} ,
\label{eq:Z_ferm}
\end{align}
where $\mathbf{k}_{\text{F}}$ is a momentum on the Fermi surface and the initial condition is 
$Z_{f1}^{\Lambda^{\rm{UV}}}=1$. 
The momentum renormalization factor is obtained from a momentum gradient of the fermion self-energy,
\begin{align}
A^{\Lambda}_{f1}
%\equiv A^{\Lambda}_{f1}[k_0 = 0, \mathbf{k}=\mathbf{k}_{\text{HS}}]
=1+\frac{|\mathbf{n}_{\mathbf{k},1}\cdot \nabla \Sigma^{\Lambda}_{f1}(k_0,\mathbf{k})|}{
|\nabla \xi_{\mathbf{k},1}|}\Big |_{k_0=0, \mathbf{k}=\mathbf{k}_{\text{F}}}\;,
\label{eq:FS_normals}
\end{align}
with the initial condition $A^{\Lambda^{\text{UV}}}_{f1}=1$. Here, $\nabla=\left(\partial_{k_x},\partial_{k_y}\right)$ and 
$\mathbf{n}_{\mathbf{k},1}$ is unit normal vector onto the Fermi surface of flavor $1$. We see 
below that the momentum gradient scales differently than the frequency derivative at the quantum critical point. 
In a different context, for Fermi systems with van Hove singularities, this asymmetry was established to all orders in perturbation theory 
by Feldman and Salmhofer.\cite{feldman08}
Necessary conditions to discover this are (i) the codimension of the Fermi surface manifold is greater than zero 
(it is zero in a one-dimensional Fermi systems) and (ii) one includes the additional, relevant transversal momentum direction parallel to the Fermi surface into the analysis.

\begin{figure}[t]
\vspace*{4mm}
\includegraphics*[width=85mm,angle=0]{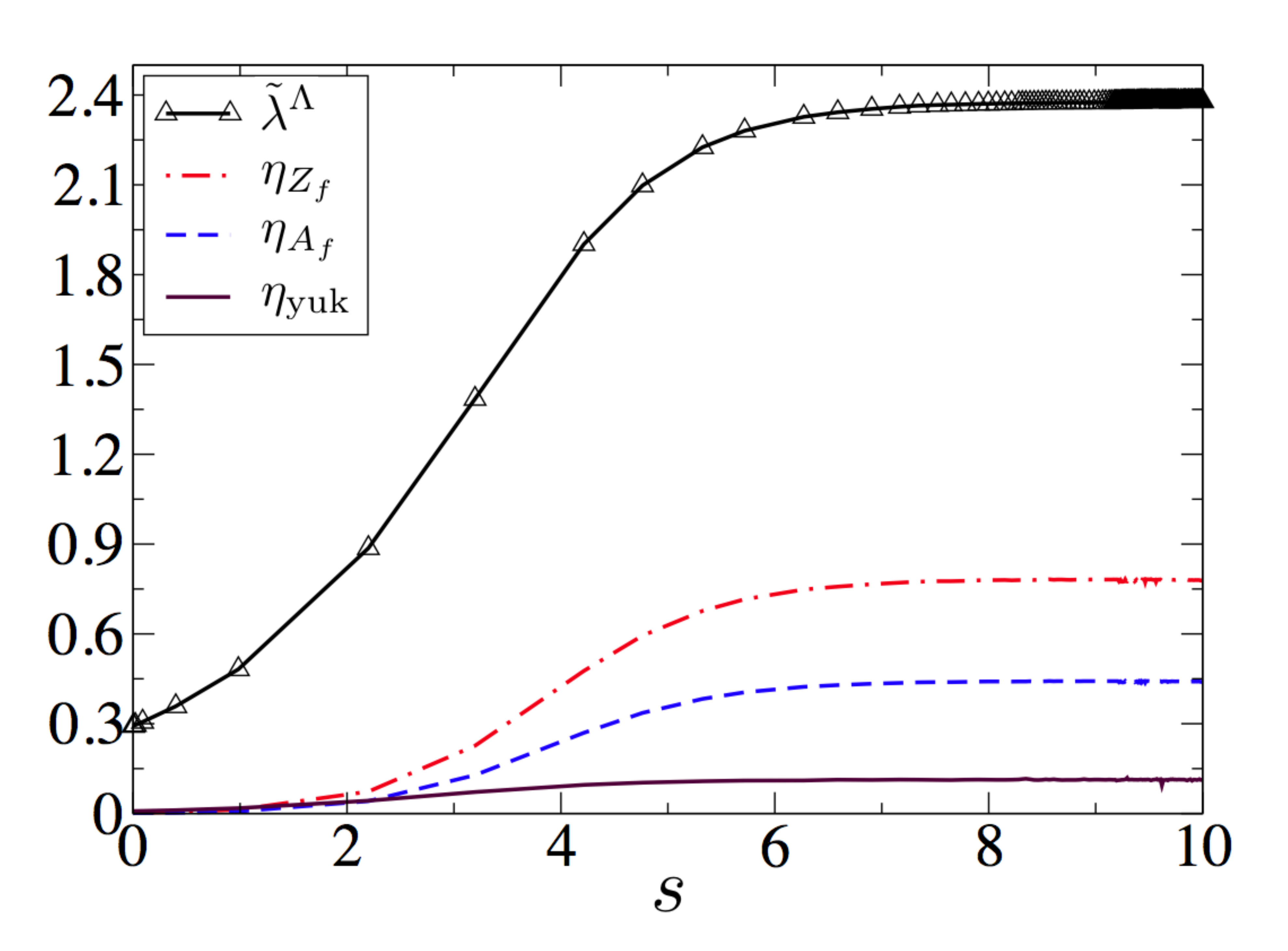}
%\includegraphics*[width=85mm,angle=0]{new_z_f.pdf}
%\vspace*{0mm}
\caption{(Color online) Quantum critical RG flows of the Yukawa coupling and the 
anomalous exponents at the hot spot $\mathbf{k}_{\text{HS}}$. The fixed-point values are 
$\tilde{\lambda}^\Lambda=2.38$, $\eta_{Z_f}=0.78$, $\eta_{A_f}=0.44$, and $\eta_{\text{yuk}}=0.11$.
The scaling plateaus for $s\gtrsim6$ depicted over $\sim$ 4 orders of magnitude would be attained 
indefinitely but are limited by the numerics only. The infrared is to the right of the plot ($\Lambda=\Lambda_{\text{UV}} e^{-s}$).}
\label{fig:new_etas_yuk}
\end{figure}

With these definitions, the scale-dependent ``dressed'' fermion propagator which occurs self-consistently in 
all RG equations becomes
\begin{align}
G^\Lambda_{f1}(k)
&=
\frac{-1}{- ik_0+\xi_{\mathbf{k},1} + \Sigma^{\Lambda}_{f1}(k_0,\mathbf{k})}
%=
%\frac{-1}{- Z^{\Lambda}_{f1}\, ik_0+A^\Lambda_{f1} \xi_{\mathbf{k},1}}
=
\frac{\mathcal{Z}^{\Lambda}_{f1}}{ik_0 - |\upsilonup^{\Lambda}_{f1}| \xi_{\mathbf{k},1}}\;,
\label{eq:ferm_prop}
\end{align}
with $\mathcal{Z}^{\Lambda}_{f1} = 1/Z_{f1}^{\Lambda}$ resembling the quasiparticle 
weight at low energies and the effective 
modulus of the Fermi velocity
%
%\begin{align}
$|\upsilonup^{\Lambda}_{f1}|=\frac{A^{\Lambda}_{f1}}{Z^{\Lambda}_{f1}}$.
%\end{align}
%

A self-consistent numerical solution of the flow equations for the Yukawa vertex $\lambda^\Lambda$, 
the fermion self-energy $\Sigma_f^\Lambda(k_0,\mathbf{k})$, and the boson 
self-energy $\Sigma_b^\Lambda(q_0,\mathbf{q})$ is attracted toward an infrared strong-coupling fixed point. 
As can be read off from Fig.~\ref{fig:new_etas_yuk}, the $\beta$ function for the Yukawa coupling [Eq.~(\ref{eq:flow_yuk_rescaled})]
vanishes for 
$s\gtrsim 6$, resulting in a scaling relation for the fermion and Yukawa anomalous exponents:
\begin{align}
\frac{d \ln \tilde{\lambda}^\Lambda}{d \ln \Lambda} = 0\quad\Leftrightarrow\quad 
\frac{1}{2} \left(\eta_{Z_f}+\eta_{A_f}\right) = \eta_{\text{yuk}} + \frac{1}{2}\;,
\label{eq:strong_coupling_FP}
\end{align}
where we dropped the flavor index as they become degenerate at the hot spot.
Similar strong-coupling fixed-point and scaling relations (without singular vertex corrections) 
have recently been obtained at the QCP of a Dirac cone toy model  
between a semimetal and a superfluid.\cite{strack10}

\begin{figure}[t] 
\vspace*{-2mm}
%\includegraphics*[width=45mm,angle=0]{Zf1_perp.pdf}
%\hspace*{-6mm}
\includegraphics*[width=90mm,angle=0]{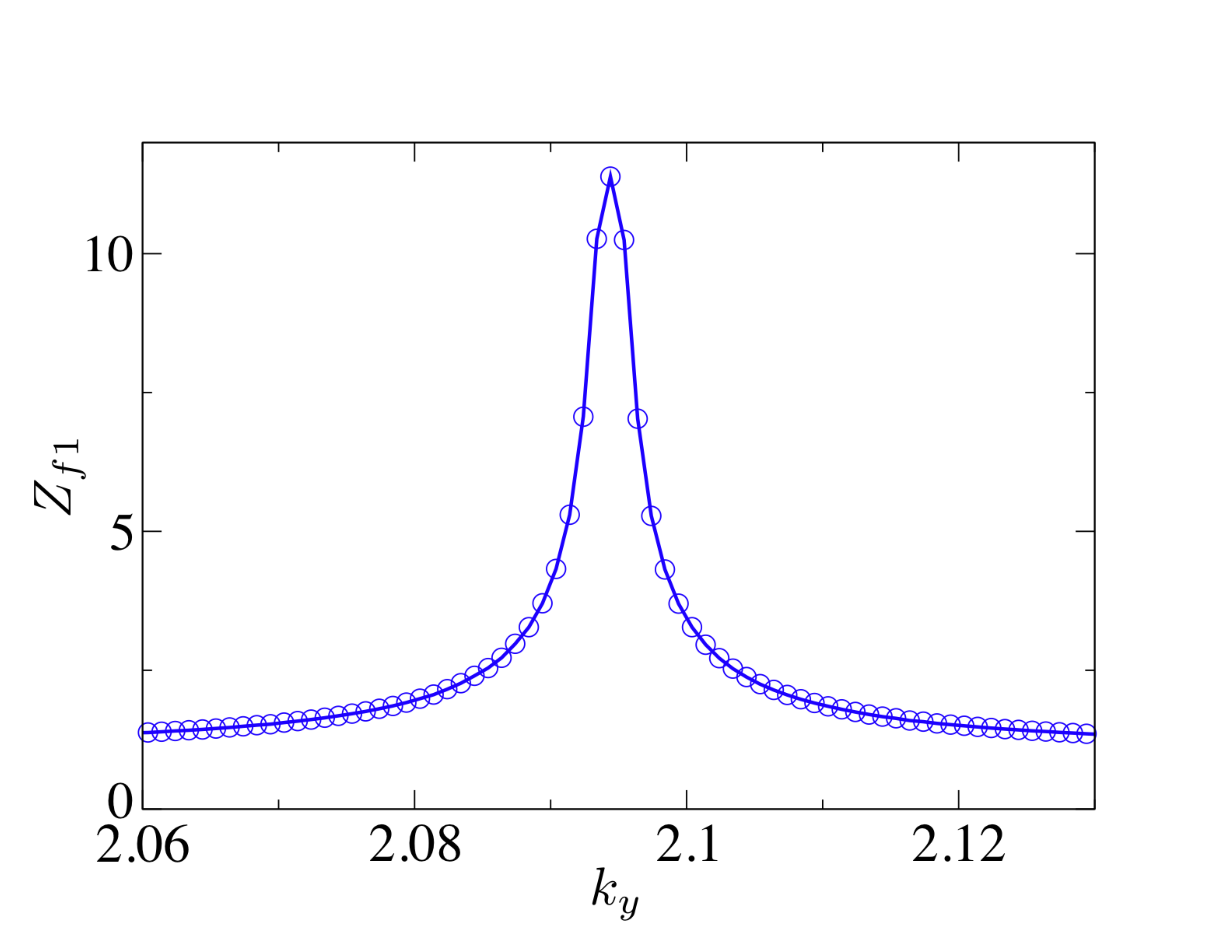}
%\hspace*{-4mm}
\caption{(Color online). Infrared values of the momentum resolved inverse quasiparticle weights
$Z^{\Lambda\rightarrow 0}_{f1}[k_0=0,k_x,k_y]$ non-self-consistently computed from Eq.~(\ref{eq:Z_ferm}) 
along the Fermi surface. Figure \ref{fig:grid_etas} exhibits flows of the corresponding exponents for the six data points 
closest to the maximum/hot spot on the right flank. Here the hot spot is located at $k_{\text{HS},y}=2.0944$ and $k_{\text{HS},x}=1.0472$.}
\label{fig:Zf_grid}
\end{figure}

\begin{figure}[t]
\vspace*{-5mm}
\includegraphics*[width=90mm,angle=0]{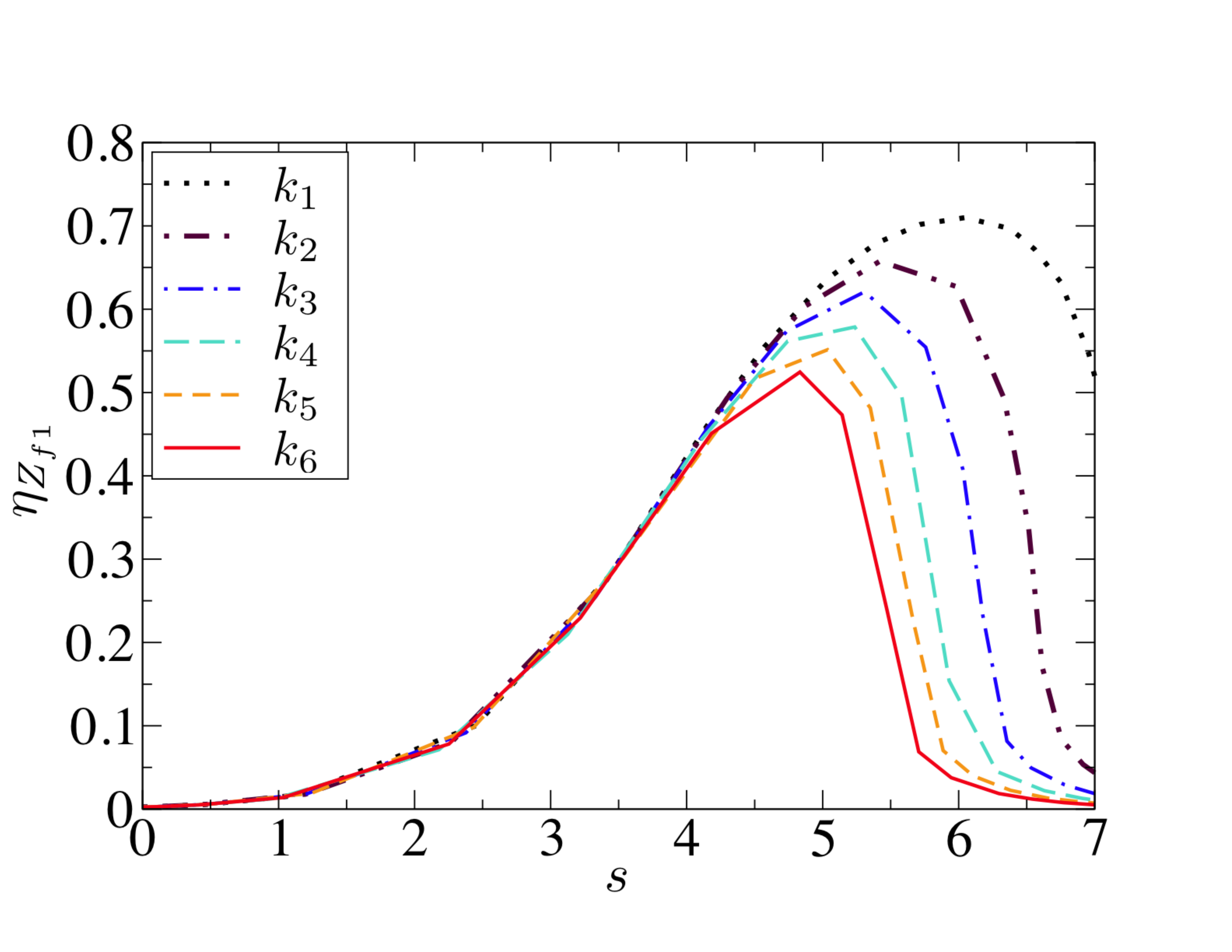}
%\vspace*{-5mm}
\caption{(Color online). Non-Fermi liquid regimes at intermediate scales of the
anomalous exponent for the quasiparticle weight $\eta_{Z_{f1}}[k_0=0,k_x,k_y]$ 
for six choices of momenta progressively approaching the hot spot (corresponding to the six data points closest to the maximum/hot spot on the right flank of Fig.~\ref{fig:Zf_grid}). 
The momentum $\mathbf{k}_6$ is furthest from 
the hot spot and $\mathbf{k}_1$ is closest to it. 
The infrared is to the right of the plot ($\Lambda=\Lambda_{\text{UV}} e^{-s}$).}
%\vspace*{-5mm}
\label{fig:grid_etas}
\end{figure}

The numerical values of the exponents (see Fig.~\ref{fig:new_etas_yuk})
%
%\begin{align}
%\eta_{Z_f}&=0.78\nonumber\\
%\eta_{A_f}&=0.44\nonumber\\
%\eta_{\text{yuk}}&=0.11\;
%\end{align}
%
determine the scaling behavior of the fermion propagator [Eq.~(\ref{eq:ferm_prop})]
and the associated dynamical exponent $z_f$. The Yukawa vertex  
diverges as a power law,
\begin{align}
 \lambda^{\Lambda\rightarrow0} \sim \frac{1}{\Lambda^{\eta_{\text{yuk}}}}=\frac{1}{\Lambda^{0.11}}\;.
\end{align}
$\Lambda$ can be associated with the momentum distance from the hot spot; 
at $\Lambda=0$ the hot spots are resonantly connected by the ordering 
wave vector $\mathbf{K}$ of the incipient
SDW. At the hot spot, the fermionic quasiparticle weight
vanishes as a power law,
\begin{align}
\mathcal{Z}^{\Lambda \rightarrow 0}_{f} \sim \Lambda^{\eta_{Zf}}=\Lambda^{0.78},
\label{eq:quasi_p}
\end{align}
destroying the Fermi liquid character of fermionic quasiparticle excitations. In a non-self-consistent calculation 
we can also compute the fermion self-energy from Eq.~(\ref{eq:Z_ferm}) away from the hot spot by solving 
the flow equations evaluated at general fermionic momenta $\mathbf{k}$. The result for a momentum cut along the
Fermi surface is exhibited in Fig.~\ref{fig:Zf_grid}. The renormalization of the quasiparticle weight is strongly peaked around the intersection of the 
Fermi surfaces at the hotspot. Away from the hot spot, the suppression of the quasiparticle weight is less pronounced, leading to asymptotically vanishing anomalous exponents in the infrared $\Lambda\rightarrow 0$ (Fig.~\ref{fig:grid_etas}). 
Nevertheless, in the vicinity of the hot spot, magnetic fluctuations are still very strong, leading to sizable non-Fermi liquid scaling regimes at intermediate scales with the maximum progressively approaching the hot-spot value $\eta_{Z_{f1}}[k_0=0,k_x=k_{\text{HS},x},k_y=k_{\text{HS},y}]=0.78$ for momenta closer to it. 

In the numerics for Fig.~\ref{fig:Zf_grid}, we stopped the flow at $s=7$ (recall that $\Lambda=\Lambda_{\text{UV}}e^{-s}$), leading 
to finite (but very large) values of $Z_{f1}$ even at the hot spot. We used a momentum cut 
of 100 points producing for each grid point in Fig.~\ref{fig:Zf_grid} the scale-resolved flows shown in Fig.~\ref{fig:grid_etas}.

The Fermi velocity vanishes as well but with a smaller exponent, 
\begin{align}
|\upsilonup^{\Lambda\rightarrow 0}_{f}|\sim \Lambda ^{\eta_{Zf}-\eta_{Af}} = \Lambda ^{z_f - 1}=\Lambda^{0.34}\;,
\label{eq:fermi_v}
\end{align}
so that the dynamical exponent for the fermions is 
\begin{align}
z_f=1+\eta_{Z_f}-\eta_{A_f}=1.34\;.
\label{eq:z_f}
\end{align}
An important ingredient to the scaling laws above is the self-consistently flowing boson propagator [Eqs.~(\ref{eq:bose_prop}) and (\ref{eq:Sigma_b})].
The asymptotic static and dynamic scaling of the spin fluctuation propagator is given by
\begin{align}
\lim_{\Lambda\rightarrow0} \left[D^{R}(q_0,\mathbf{q})\right]^{-1}
&\sim \Lambda^{\eta_{Z_b}}  |q_0| +  \mathbf{q}^2 
\sim
|q_0|^{1.66} + \mathbf{q}^2\;,
\end{align}
with $\eta_{Z_b}=0.66$. Remarkably, the boson dynamical exponent,
\begin{align}
z_b = 2 - \eta_{Z_b} = 1.34 = z_f\;,
\label{eq:z_b}
\end{align}
takes the same value as the fermion dynamical exponent. It is a distinguishing feature of this 
infrared fixed-point of electrons in metals at a SDW transition
that the dynamical exponent attains fractional value different from $1$ (which is the exact value 
for quantum-critical fermion systems with Lorentz-symmetry; see Ref.~\onlinecite{janssen12} and references therein) and 
different from $2$ (which is the mean-field value of the Hertz theory \cite{hertz}). Our 
fermion anomalous dimensions and $z$ can be mapped 
to those of Ref.~\onlinecite{metlitski10-2} for values of the Fermi velocity-anisotropy in a range around 
$\alpha \approx 0.5$ and upon ignoring the marginal
RG flow of $\alpha$ [which is implicitly assumed in (\ref{eq:FS_normals})]; our boson anomalous dimension renormalizing 
the $\mathbf{q}^2$ term in the propagator is essentially zero,
and we trace this to differences in the RG scheme from Ref.~\onlinecite{metlitski10-2}.

\section{Conclusion\label{conc}}

This paper was dedicated to the critical behavior of compressible, electronic quantum matter in two-dimensional lattices interacting with self-generated, singular antiferromagnetic fluctuations. We generalized previous hot-spot theories to full ``UV-completed'' Fermi surfaces free of spurious edge singularities in a model that can also be analyzed with quantum Monte Carlo. This should enable a cross-fertilizing comparison of results obtained with different methods for this problem. We provided first quantitative estimates for the critical exponents of the single-particle and spin fluctuations correlators which deviate strongly from the Hertz-Millis values. The solution of our RG equations was attracted toward a stable, strong-coupling fixed point, resulting in a common dynamic exponent for the fermions and the bosons.

It would be interesting to classify all relevant operators to our fixed point and investigate 
the stability of our strong-coupling fixed point further. As a first simple step in this direction, we have extended the truncation for the fermion dispersions to allow for changes in the Fermi surface curvature (keeping the position of the hot spot fixed). A scale-dependent $\tilde{\alpha}^\Lambda$ that modifies the hoppings, $ t_{1,x/y}\rightarrow t_{1,x/y} + \tilde{\alpha}^\Lambda$ and 
$t_{2,x/y}\rightarrow t_{2,x/y} - \tilde{\alpha}^\Lambda$, does the job. We found only relatively small, finite renormalizations of $\tilde{\alpha}^\Lambda$. 
However, a proper self-consistent investigation of a flowing Fermi surface with the full dispersion used in this paper requires an advanced truncation and likely also a self-consistent determination also of the position of the Fermi surfaces and the hot spots as a function of $\Lambda$. Potential tendencies toward magnetic ordering at incommensurate wave vectors might also be captured that way. 
Such a state-of-the-art truncation was recently presented for self-energy flows in the repulsive Hubbard model close to van Hove filling.\cite{giering12}

Other promising future directions are the inclusion of ($d$-wave) superconductivity,\cite{sedeki12} an extension to the quantum-critical regime at finite temperatures, and the exploration of the antiferromagnetic phase with broken symmetry close to the quantum critical point.\footnote{For example, by generalizing Ref.~\onlinecite{strack08} from the superfluid $O(2)$ case to the staggered $O(3)$ case for the spin-fermion model.}

\begin{acknowledgements}
We thank E. Berg, C. Honerkamp, E. G. Moon, and M. Punk for useful discussions. This research was supported by the DFG under grant Str 1176/1-1, by the NSF under Grant DMR-1103860, and by 
the Army Research Office Award W911NF-12-1-0227. JL is also supported by the STX Foundation.
\end{acknowledgements}

\begin{widetext}

\appendix*
\section{Explicit form of flow equations}

%\begin{widetext}

We here give the explicit expressions of the flow equations (\ref{eq:flow_ferm}), (\ref{eq:flow_bos}), (\ref{eq:flow_yuk}). To that end, it is 
convenient to use the rescaled variables $\tilde{Z}^{\Lambda}_{b}=\frac{Z_b^\Lambda}{\Lambda}$, 
$\tilde{\xi}_{\mathbf{k},1}=\frac{\xi_{\mathbf{k},1}}{\Lambda}$ as well as 
rescaled momenta: $\tilde{k}_0=\frac{k_0}{\Lambda}$, $\tilde{q}_0=\frac{q_0}{\Lambda}$, 
$\tilde{q}_x=\frac{q_x}{\Lambda}$, and $\tilde{q}_y=\frac{q_y}{\Lambda}$.

For the the fermionic frequency exponent, there is
%
%\begin{widetext}
%
\begin{align}
\eta_{Z_{f1}}=3 \left(\tilde{\lambda}^\Lambda\right)^2
\sqrt{|\upsilonup^{\Lambda}_{f1}|\, |\upsilonup^{\Lambda}_{f2}|}
\int^1_{-1}\frac{ d \tilde{q}_y} {2\pi}
\int_{-\sqrt{1-\tilde{q}_y^2}}^{+\sqrt{1-\tilde{q}_y^2}}
\frac{ d \tilde{q}_x} {2\pi}
\int^{\infty}_{-\infty}\frac{ d \tilde{q}_0}{2\pi}
2 A^{\Lambda}_b 
\frac{1}{\left( i \tilde{q}_0 - |\upsilonup^{\Lambda}_{f2}| \tilde{\xi}_{\mathbf{k}_{\rm{HS}}+\tilde{\mathbf{q}},2}\right)^2}
\frac{1}{\left(\tilde{Z}^\Lambda_b |\tilde{q}_0| + A^\Lambda_b\right)^2}\;,
\label{eq:eta_Zf1}
\end{align}
and similarly ($1\leftrightarrow 2$) for flavor 2. The frequency integral over $\tilde{q}_0$ can be performed analytically so that 
at each step of the flow, two-dimensional integrations over $\tilde{q}_x$ and $\tilde{q}_y$ have to be performed numerically. 
The Yukawa anomalous exponent contains fermion propagators of both flavors:
\begin{align}
\eta_{\rm{yuk}}= - 
\left(\tilde{\lambda}^\Lambda\right)^2
\sqrt{|\upsilonup^{\Lambda}_{f1}|\, |\upsilonup^{\Lambda}_{f2}|}
\int^1_{-1}\frac{ d \tilde{q}_y} {2\pi}
\int_{-\sqrt{1-\tilde{q}_y^2}}^{+\sqrt{1-\tilde{q}_y^2}}
\frac{ d \tilde{q}_x} {2\pi}
\int^{\infty}_{-\infty}\frac{ d \tilde{q}_0}{2\pi}
2 A^{\Lambda}_b 
\frac{1}{i \tilde{q}_0 -|\upsilonup^{\Lambda}_{f1}| \tilde{\xi}_{\mathbf{k}_{\rm{HS}}+\tilde{\mathbf{q}},1}}
\frac{1}{i \tilde{q}_0 -|\upsilonup^{\Lambda}_{f2}| \tilde{\xi}_{\mathbf{k}_{\rm{HS}}+\tilde{\mathbf{q}},2}}
\frac{1}{\left(\tilde{Z}^\Lambda_b |\tilde{q}_0| + A^\Lambda_b\right)^2}\;.
\label{eq:eta_yuk}
\end{align}
%\end{widetext}
%
For the flow of the fermionic momentum factors 
we use the projected $k_x$ and $k_y$ derivatives of Eq.~(\ref{eq:flow_ferm}),
\begin{align}
\partial_\Lambda A^{\Lambda}_{f1,x}&=
%\frac{1}{2t_{1,x}\sin(k_x)}
n_{k_{x},1} \partial_{k_x}\partial_{\Lambda}\Sigma_{f1}^{\Lambda}[k_0,\mathbf{k}]\Big|_{k_0=0,\mathbf{k}=\mathbf{k}_{\text{HS}}},
\nonumber\\
\partial_\Lambda A^{\Lambda}_{f1,y}&=
%\frac{1}{2t_{1,y}\sin(k_y)}
n_{k_{y},1} \partial_{k_y}\partial_{\Lambda}\Sigma_{f1}^{\Lambda}[k_0,\mathbf{k}]\Big|_{k_0=0,\mathbf{k}=\mathbf{k}_{\text{HS}}},
\end{align}
with the initial conditions $A_{f1,x}^{\Lambda^{\rm{UV}}}=A_{f1,y}^{\Lambda^{\rm{UV}}}=1$. 
The Fermi surface normal projector is (similarly for flavor 2)
\begin{align}
n_{k_{x/y},1}=\frac{2 t_{1,x/y} \sin k_{x/y}}{\sqrt{\left(2 t_{1,x} \sin k_{x/y}\right)^2 +  \left(2 t_{1,y} \sin k_y\right)^2}}\;.
%n_{k_y,1}&=\frac{2 t_{1,y} \sin k_y}{\sqrt{\left(2 t_{1,x} \sin k_x\right)^2 +  \left(2 t_{1,y} \sin k_y\right)^2}}\;,
\label{eq:n_xy}
\end{align}
The flow equations 
for the rescaled variables are
%
%\begin{align}
$\tilde{A}^{\Lambda}_{f1,x}=\frac{A^{\Lambda}_{f1,x}}{Z^\Lambda_{f1}}$, 
$\tilde{A}^{\Lambda}_{f1,y}=\frac{A^{\Lambda}_{f1,y}}{Z^\Lambda_{f1}}$.
%\end{align}
%
With $\eta_{Z_{f1}}$ given in Eq.~(\ref{eq:eta_Zf1}), these take the form
\begin{align}
\Lambda\partial_{\Lambda} \tilde{A}^{\Lambda}_{f1,x} &= \left(\eta_{Z_{f1}}-\eta_{A_{f1,x}}\right)\tilde{A}^{\Lambda}_{f1,x},
\nonumber\\
\Lambda\partial_{\Lambda} \tilde{A}^{\Lambda}_{f1,y} &= \left(\eta_{Z_{f1}}-\eta_{A_{f1,y}}\right)\tilde{A}^{\Lambda}_{f1,y}\;,
\label{eq:Afxy_tilde}
\end{align}
with the exponents
%
%\begin{align}
$\eta_{A_{f1,x}}=-\frac{ d \ln A^\Lambda_{f1,x}}{d \ln \Lambda}$, %\;,\quad
$\eta_{A_{f1,y}}=-\frac{ d \ln A^\Lambda_{f1,y}}{d \ln \Lambda}$.%\;,\quad
%\end{align}
%
At every step of the flow, we compute then per Eq.~(6) %\ref{eq:FS_normals}
\begin{align}
|\upsilonup^{\Lambda}_{f1}|=
\frac{\sqrt{\left(\tilde{A}^\Lambda_{f1,x}\right)^2 
+ \left( \tilde{A}^\Lambda_{f1,y}\right)^2 }}
{|\nabla \xi_{1,\mathbf{k}}|_{\mathbf{k}=\mathbf{k}_{\rm HS}}}\;.
\label{eq:v_fermi}
\end{align}
Expressions for the exponents:
%\begin{widetext}
\begin{align}
\eta_{A_{f1,x}}&=-n_{k_{\rm{HS},x},1}\, 3 \left(\tilde{\lambda}^\Lambda\right)^2
\sqrt{|\upsilonup^{\Lambda}_{f1}|\, |\upsilonup^{\Lambda}_{f2}|}
\frac{ |\upsilonup^{\Lambda}_{f2}|}{\tilde{A}_{f1,x}}
\int^1_{-1}\frac{ d \tilde{q}_y} {2\pi}
\int_{-\sqrt{1-\tilde{q}_y^2}}^{+\sqrt{1-\tilde{q}_y^2}}
\frac{ d \tilde{q}_x} {2\pi}
\int^{\infty}_{-\infty}\frac{ d \tilde{q}_0}{2\pi}
2 A^{\Lambda}_b 
\frac{2 t_{2x} \sin\left(k_{\rm{HS},x}+\tilde{q}_x \Lambda\right)}{\left( i \tilde{q}_0 - |\upsilonup^{\Lambda}_{f2}| \tilde{\xi}_{\mathbf{k}_{\rm{HS}}+\tilde{\mathbf{q}},2}\right)^2}
\frac{1}{\left(\tilde{Z}^\Lambda_b |\tilde{q}_0| + A^\Lambda_b\right)^2},
\nonumber\\
\eta_{A_{f1,y}}&=-n_{k_{\rm{HS},y},1}\,3 \left(\tilde{\lambda}^\Lambda\right)^2
\sqrt{|\upsilonup^{\Lambda}_{f1}|\, |\upsilonup^{\Lambda}_{f2}|}
\frac{ |\upsilonup^{\Lambda}_{f1}|}{\tilde{A}_{f1,y}}
\int^1_{-1}\frac{ d \tilde{q}_y} {2\pi}
\int_{-\sqrt{1-\tilde{q}_y^2}}^{+\sqrt{1-\tilde{q}_y^2}}
\frac{ d \tilde{q}_x} {2\pi}
\int^{\infty}_{-\infty}\frac{ d \tilde{q}_0}{2\pi}
2 A^{\Lambda}_b 
\frac{2 t_{2y} \sin\left(k_{\rm{HS},y}+\tilde{q}_y \Lambda\right)}{\left( i \tilde{q}_0 - |\upsilonup^{\Lambda}_{f2}| \tilde{\xi}_{\mathbf{k}_{\rm{HS}}+\tilde{\mathbf{q}},2}\right)^2}
\frac{1}{\left(\tilde{Z}^\Lambda_b |\tilde{q}_0| + A^\Lambda_b\right)^2}\;.
\label{eq:eta_Afxy}
\end{align}
%\end{widetext}
%

%\begin{widetext}
Finally, the (rescaled) boson frequency factor and momentum factor are self-consistently determined from
%
%\begin{widetext}
\begin{align}
\tilde{Z}^\Lambda_{b}&= 2 \left(\tilde{\lambda}^{\Lambda}\right)^2\sqrt{|\upsilonup^{\Lambda}_{f1}|\, |\upsilonup^{\Lambda}_{f2}|}
\int^\pi_{-\pi} \frac{d k_x}{2\pi} \int^\pi_{-\pi} \frac{d k_y}{2\pi} 
\frac{1}{\Lambda^2}
\int ^\infty_{-\infty} \frac{ d \tilde{k}_0}{2\pi} 
\left[
\left(
\frac{1}{i(\tilde{k}_0+1) - |\upsilonup^{\Lambda}_{f1}| \xi_{\mathbf{k},1}} - \frac{1}{i\tilde{k}_0-
|\upsilonup^{\Lambda}_{f1}| \xi_{\mathbf{k},1}}
\right)
 \frac{1}{i\tilde{k}_0-
|\upsilonup^{\Lambda}_{f2}| \xi_{\mathbf{k},2}}
+ \left(1 \leftrightarrow 2\right)
\right],
\nonumber\\
\tilde{A}^\Lambda_{b}&= 2 \left(\tilde{\lambda}^{\Lambda}\right)^2\sqrt{|\upsilonup^{\Lambda}_{f1}|\, |\upsilonup^{\Lambda}_{f2}|}
\int^\pi_{-\pi} \frac{d k_x}{2\pi} \int^\pi_{-\pi} \frac{d k_y}{2\pi} 
\frac{1}{\Lambda^2}
\int ^\infty_{-\infty} \frac{ d \tilde{k}_0}{2\pi} 
\left[
\left(
\frac{1}{i\tilde{k}_0 - |\upsilonup^{\Lambda}_{f1}| \xi_{\mathbf{k}+q_x,1}} - \frac{1}{i\tilde{k}_0-
|\upsilonup^{\Lambda}_{f1}| \xi_{\mathbf{k},1}}
\right)
 \frac{1}{i\tilde{k}_0-
|\upsilonup^{\Lambda}_{f2}| \xi_{\mathbf{k},2}}
+ \left(1 \leftrightarrow 2\right)
\right]_{q_x=\Lambda}\;.
\label{eq:ZA_b}
\end{align}
%\end{widetext}
%

Equations (\ref{eq:flow_yuk_rescaled}), (\ref{eq:eta_Zf1}), (\ref{eq:eta_yuk}), and (\ref{eq:Afxy_tilde})--(\ref{eq:ZA_b}) are solved numerically as a function of flow parameter $\Lambda=\Lambda^{\rm{UV}}e^{-s}$ so that $s=0$ corresponds to the UV ($\Lambda^{\rm{UV}}=1$). The hot spot coordinates are $k_{\rm{HS},x}=1.0472$, $k_{\rm{HS},y}=2.0944$.
As initial conditions, we choose $\lambda^{\Lambda_{\text{UV}}}=0.25$, $Z^{\Lambda_{\text{UV}}}_{f1}=Z^{\Lambda_{\text{UV}}}_{f2}=1$, and $A^{\Lambda_{\text{UV}}}_{f1}=A^{\Lambda_{\text{UV}}}_{f2}=1$. The initial values 
for the boson propagator are $\tilde{Z}_{b}^{\Lambda_{\text{UV}}}=0.052$ and $\tilde{A}_{b}^{\Lambda_{\text{UV}}}=1.011$. 

\end{widetext}

\end{document}